\def\year{2019}\relax
\documentclass[letterpaper]{article} 

\usepackage{aaai19}  
\usepackage{times}  
\usepackage{helvet}  
\usepackage{courier}  
\usepackage{url}  
\usepackage{graphicx}  
\usepackage{url}
\makeatletter
\g@addto@macro{\UrlBreaks}{\UrlOrds}
\makeatother

\usepackage{enumerate}

\usepackage{booktabs}

\usepackage{tabularx}

\usepackage{amsmath}

\usepackage{subcaption}
\newcolumntype{Z}{>{\centering\let\newline\\\arraybackslash\hspace{0pt}}X}

\makeatletter
\def\copyright@on{F}
\def\copyright@text{\relax}
\makeatother

\begin{document}
%
\title{Health and Kinship Matter: Learning About Direct-To-Consumer Genetic Testing User Experiences via Online Discussions}

\author{Zhijun Yin, Lijun Song, Ellen Clayton and Bradley Malin\\
 Vanderbilt University, Nashville, TN, USA\\
 {zhijun.yin,lijun.song,ellen.clayton,b.malin}@vanderbilt.edu\\
 }

\maketitle
\begin{abstract}
Direct-to-consumer (DTC) genetic testing has gained in popularity over the past decade, with over 12 million consumers to date. Given its increasing stature in society, along with weak regulatory oversight, it is important to learn about actual consumers' testing experiences. Traditional interviews or survey-based studies have been limited in that they had small sample sizes or lacked detailed descriptions of personal experiences. Yet many people are now sharing their DTC genetic testing experiences via online social media platforms. In this paper, we focused on one particularly lively online discussion forum, \textit{r/23andme} subreddit, where, as of before March 2018, 5,857 users published 37,183 posts. We applied topic modeling to the posts and examined the identified topics and temporal posting trends. We further applied regression analysis to learn the association between the attention that a submission received, in terms of votes and comments, and the posting content. Our findings indicate that bursts of the increase of such online discussion in 2017 may correlate with the Food and Drug Administration's authorization for marketing of 23andMe genetic test on health risks, as well as hot sale of 23andMe's products on Black Friday. While ancestry composition was a popular subject, kinship was steadily growing towards a major online discussion topic. Moreover, compared with other topics, health and kinship were more likely to receive attention, in terms of votes, while testing reports were more likely to receive attention, in term of comments. Our findings suggest that people may not always be prepared to deal with the unexpected consequences of DTC genetic testing. Moreover, it appears that the users in this subreddit might not sufficiently consider privacy when taking a test or seeking an interpretation from a third-party service provider.
\end{abstract}

\section{Introduction}

Genetic testing can be applied to determine ancestry, kinship, lifestyle and the risk of developing common diseases (e.g., Parkinson disease and Alzheimer disease). Traditionally, these tests have been ordered and interpreted by healthcare providers or biomedical researchers. However, the past decade has witnessed a growing popularity in direct-to-consumer (DTC) genetic testing. DTC genetic testing, like any ordinary product in our daily life, can now be ordered by consumers themselves from various companies, such as Ancestry.com or 23andMe.com. Reports suggest that the total number of people who have already participated in DTC genetic testing increased from 4.5 million to 12 million in 2017 \cite{regalado2018}. 

Given such a large, and growing number of participants, it is important to understand people's motivation and experiences with engaging with a DTC genetic testing service, processing the test results, and handling the possible consequences. This is because, in spite of the broad benefits brought by DTC genetic testing, such as the promotion of awareness of genetic diseases and personalized genetic information, there still exists a number of risks and limitations. These include a lack of understanding of testing results, as well as the generation of unexpected information about health and family relationships. Additionally, individuals are contributing their records to various online sites, such as GEDmatch (a free online database), which have provided society benefit by the discovery of criminals associated with cold cases, such as the Golden State Killer \cite{erlich2018identity}. Yet, at the same time, such resources could be used to potentially identify millions of Americans - even if they never underwent DTC genetic testing or consented to the sharing of information about themselves - leading to concerns about privacy intrusion \cite{ducharme2018}.

There is a growing body of survey-based studies that have focused on investigating the public's attitude, as well as knowledge about, DTC genetic testing. However, a systemic literature review, which summarized over one hundred of such studies, found that only nine investigated the experience of participants who actually purchased the service \cite{covolo2015internet}. While learning about the general population's awareness or attitude can provide insights into knowledge gaps and the possibility of widening health-related disparities \cite{salloum2018rural}, examining actual consumers' experiences and feelings can lead to a better understanding of the personal impact of DTC genetic testing. As an alternative, it is possible to interview actual consumers to obtain more detailed information, but this approach is limited in that it is quite time-consuming and is generally not scalable in sample size \cite{yin2017power}. 

At the same time, many people now use online environments to discuss and share many aspects of their daily life, including their DTC genetic test results. For instance, it was shown that Twitter users often post their ethnic background more than other information, such as disease risks, with respect to testing results  \cite{olejnik2014m}. More recently, a large-scale analysis of Twitter discourse related to DTC genetic testing indicated that this behavior was often influenced by news and DTC websites \cite {mittos201823andme}. Though there is a large population on Twitter who have disclosed their test results, it is difficult for Twitter users to develop rich discussions due to the limited number of characters in a tweet and its design as an all-purpose discussion environment.  This substantially limits its applicability for gaining a deeper understanding of an individual's attitude about, and experience with, DTC genetic testing.

In this study, we investigate personal discussions of DTC genetic testing on Reddit, an online content rating and discussion website. Unlike Twitter, which maintains its content based on a social network, Reddit organizes its content into different subreddits based on topics  (e.g., \textit{r/legaladvice} and \textit{r/gaming}) and imposes no limitations on posting length. In each subreddit, users can initiate a new thread by publishing a submission post, or make comments, upvote or downvote on either submissions or comments. We focused on the data in the \textit{r/23andme} subreddit\footnote{We did not choose other subreddits (e.g., \textit{r/AncestryDNA}) because \textit{r/23andme} has much more active subscribers.}, where the users are mainly consumers of 23andMe. In \textit{r/23andme}, users are discussing a broad range of topics regarding DTC genetic testing, including testing services, testing results, interpretation and stories after taking the test (e.g., about health and kinship). 

We aimed to investigate three research questions: RQ1) What do people talk about in this subreddit? RQ2) How do the topics change over time? and RQ3) What kinds of submissions are more likely to receive attention? To do so, we adopted an approach that combined topic modeling, hierarchical clustering on words and the extraction of linguistic features to deeply examine the content of these online discussions. We analyzed the temporal trends of topics, which we compared with the posting temporal trend. Finally, we applied regression analysis to examine the association between the attention received by submissions, in term of the received comments and the karma score (the number of upvotes minus the number of downvotes), and the submission content. The main contributions of this work are as follows:

\begin{itemize}
\item We identified 12 topics (see Table~\ref{tab:topic_terms}) that were discussed in \textit{r/23andme} subreddit, which coalesced into four categories: i) kinship; ii) testing reports; iii) discussion about health and traits; and iv) the progress of testing service.

\item We found that there was a rapid increase in the number of posts per month in 2017. The bursts of the increased number of posts coincided with increase in the discussion about progress in testing. 

\item While ancestry composition was a popular subject, kinship was steadily growing towards a major online discussion topic in 2017.

\item Submissions that mentioned health or kinship were more likely to receive attention in terms of karma score. By contrast, submissions that mentioned test reports were more likely to receive attention in terms of the number of comments.
 
\end{itemize}
\section{Background and Motivations}
\label{sec:back}

In this section, we summarize the related studies, their limitations and potential opportunities to motivate our research. 

\subsection{Detecting Health Risks} 
One potential benefit of taking a genetic test is to detect health risks. For example, an interview-based study of consumers of the 23andMe BRCA test indicated that the unexpected health risks benefited participants as well as drove relatives of identified carriers to seek testing, leading more carriers identified \cite{francke2013dealing}. However, the testing results may not be accurate. It was reported that a patient who underwent DTC BRCA genetic testing received a different result from that was done in a clinic \cite{schleit2018first}. This suggests that DTC genetic testing may cause confusion or misunderstanding of result interpretation, in spite of its potential in detecting health risks. Additionally, a recent viewpoint expressed concerns about the unknown harms of overtesting after the U.S. Food and Drug Administration (FDA) granted marketing authorization to 23andMe \cite{gill2018direct}.

\subsection{Influence on Behaviors} 
Genetic testing results may influence people's behaviors, but the observed effects have been mixed. For instance, \cite{roberts2017direct} examined a survey of 1,648 consumers from 23andMe and Pathway in 2012 and found that, before taking the test, most people were interested in ancestry, trait information and disease risks. After the test, 59\% of respondents claimed that the test results would influence their health management and 2\% reported regret in taking the test. By contrast, an earlier survey-based study of 3,639 subjects showed that there were no measurable short-term changes in psychological health, diet or exercise behavior after taking the test \cite{bloss2011effect}. However, in a survey of 961 participants, it was shown that an atypical drug response was common with DTC genetic testing and was associated with prescription medication changes \cite{carere2017prescription}. Meanwhile, a genetic study of nicotine dependence found that smoking cessation attempts statistically increased after the survey participants received the test results \cite{hartz2015return}.   

\subsection{Privacy Concerns} 
Many studies have investigated the privacy concerns and ethical implications of DTC genetic testing. For example, a recent survey examined the privacy policies of ninety DTC genetic testing companies and concluded that most were not aligned with privacy frameworks endorsed by the Federal Trade Commission \cite{hazel2018knows}. Furthermore, consumers who send their testing results for a third-party interpretation make the privacy issues yet more complicated \cite{badalato2017third}. We refer readers to an systemic literature review for further details about individuals' perspectives on privacy and genetic information \cite{clayton2018systematic}.

\subsection{Social Computing}
User generated content in online environments has proven helpful for researchers to investigate a broad range of topics \cite{mejova2016fetishizing,de2016discovering}. However, only a few studies have focused on mining the online disclosure of DTC genetic testing results in Twitter \cite{olejnik2014m,mittos201823andme}. Moreover, besides the aforementioned research topics, few studies investigated the extent to which actual consumers applied and discussed their experiences regarding kinship, another popular application of DTC genetic testing. By contrast, our study focuses on the online discussion of DTC genetic testing on \textit{r/23andme} subreddit. While we acknowledge this is a specific population (e.g.  consumers of the 23andMe testing service), this is a rich environment of natural discussion, which we believe can provide greater insights into personal experiences and discussion about DTC genetic testing.
\section{Methods}

In this section, we describe the methods that we applied to investigate the three posited research questions: data collection, topic extraction, trend and regression analysis.

\subsection{Data Preparation}
We applied PRAW (version 5.6.0), a python wrapper of the Reddit official API, to collect all of the posts that were published in \textit{r/23andme} subreddit before March 22, 2018. Each post (submission or comment) contained the following fields: i) post id, ii) author name, iii) creation date, iv) title, v) body text, vi) the number of upvotes, vii) the number of downvotes, and viii) the post ID that it replied to. While each submission had a full list of related comments, we only counted the direct replies for calculating the number of comments that it received. We followed \cite{de2014mental} to calculate the karma score for each submission by subtracting the number of upvotes by the number of downvotes. We combined the title and the body text of each submission together to represent their content. This was done because we did not collect the images or text linked by URLs and some submissions might only contain URLs in the body text. We believe the titles can provide additional information about the topics of these submissions. We further removed the posts that only contained a \textit{[delete]} in the content for marking the status of being deleted.

\subsection{Profiling Discussion Content}
Due to the sparsity of natural language, it is common to summarize large quantities of free text. There are two common methods for doing so: word clustering based on word embedding techniques (e.g., word2vec) and topic modeling. While word clustering groups words that are close in either document position or semantic space, topic modeling groups words that appear in a similar global context. In this study, we relied on both of these methods to obtain the local word semantic clusters, as well as the global topics. While the global topics were applied to demonstrate the general picture of online discussions and their overall temporal trends, the word semantic clusters were applied to regression analysis because they can simultaneously provide greater details in context while maintaining interpretability.

We applied an implementation of Latent Dirichlet Allocation (LDA) \cite{blei2003latent}, a classical topic modeling technique, in Mallet (version 2.0.8) to identify the main themes of online discussions in \textit{r/23andme}. Since LDA is an unsupervised technique, we relied on coherence score to determine the optimal number of topics \cite{roder2015exploring}. Specifically, we ran LDA models for 2 to 25 topics (with step size of 1) on all of the posts and chose the number of topics with the highest coherence score. To mitigate word sparsity and ensure interpretability, we replaced each word with its lemma form and retained only nouns, verbs, adjectives, and adverbs.

To obtain the word semantic clusters, we relied on the Google pretrained word2vec model. This was done because our dataset was not sufficiently large to fit an accurate word2vec model. We adopted the same method in \cite{yin2018therapy} to determine the optimal number of clusters. Specifically, we ran the cluster algorithm with 25 to 1000 clusters (with a step size of 25) and applied the elbow rule to the standard deviation of cluster size. We selected the number of clusters at the angle where the marginal gain begins to diminish, using a heuristic that a large cluster is more likely to contain multiple topics while a small cluster is more likely to have little contribution to dimension reduction.

Additionally, we applied LIWC (version 2015) to generate linguistic features to help profile post content to account for the fact that some categories may not be represented by either topics or word semantic clusters. It should be noted that linguistic features have proven useful in social media content analysis \cite{de2016discovering}. In this study, we excluded categories of \textit{informal language} and \textit{summary languages variables} and included the other categories into our regression analysis. If a category has sub-categories (e.g., \textit{negative emotions} include \textit{anxiety}, \textit{anger} and \textit{sadness}), we only applied its sub-categories. 

\subsection{Temporal Trend Analysis}
We applied a rolling average, with time windows of 30 days and 3 months, to obtain the overall trajectory of posting frequency and topic prevalence, respectively. We counted the number of submissions and comments published in a 30-day rolling window. We followed the method proposed by \cite{guille2016tom} to compute topic prevalence. Specifically, given a 3-month rolling window, we define the prevalence of a topic as the proportion of the submissions in which this topic has the highest distribution.

\subsection{Regression Analysis}
Since users can comment or vote on a submission, we applied the number of comments and karma score (the number of upvotes minus the number of downvotes) to characterize its attention. These numbers are non-negative, such that we applied negative binomial regression to learn the association between the attention of a submissions and its content in terms of topics, word semantic clusters, and linguistic features. We applied the topic distribution and the proportion of each linguistic feature as feature values. We calculated term-frequency-inverse-document-frequency (TF-IDF) for word semantic clusters in each submission as proposed in \cite{yin2018therapy}. All of these features were first normalized and then scaled into a range of [0, 1]. We further applied the \textit{findCorrelation} function, as implemented in the caret R package (version 6.0-81) with a cutoff of 0.45 to remove correlated features. By doing so, we ensured the regression models could converge. 

It is possible that a submission that was published earlier (or with more words) might be exposed to more readers (or attractive), thus, being more likely to receive votes and comments. As such, we incorporated the post length and the live time of a submission as additional control variables when fitting models. We applied the implementation of negative binomial regression in MASS R package (version 7.3-45) to fit two models for karma score and the number of comments, respectively. Finally, we reported the features at a statistical significance level of 0.0001. 
\section{Results}

In this section, we report our results regarding the three proposed research questions.

\begin{figure*}[ht]
\centering
        \begin{subfigure}[b]{0.33\textwidth}
                \includegraphics[width=\linewidth]{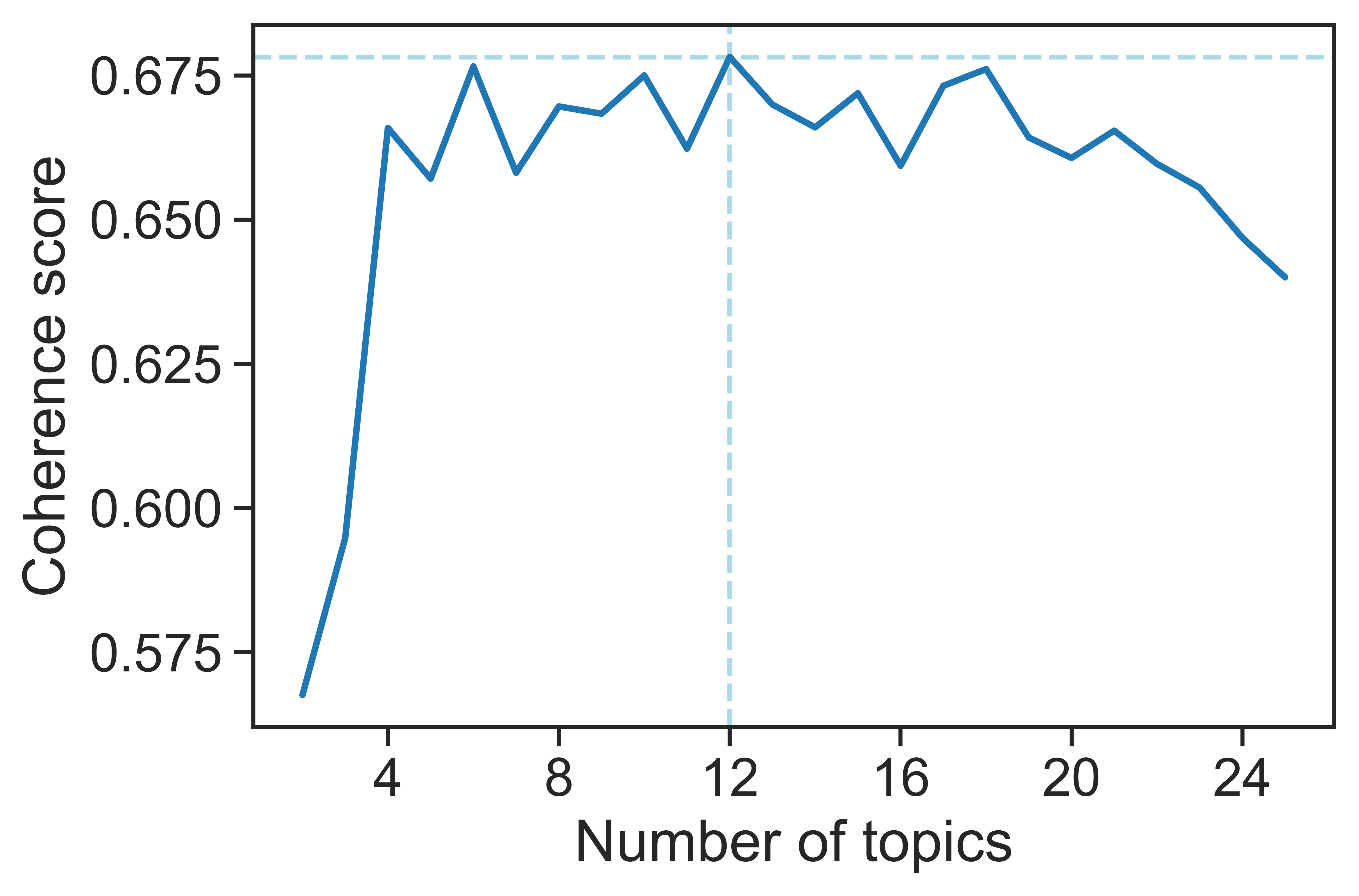}
                \caption{}
                \label{fig:coherence}
        \end{subfigure}%
        \begin{subfigure}[b]{0.33\textwidth}
                \includegraphics[width=\linewidth]{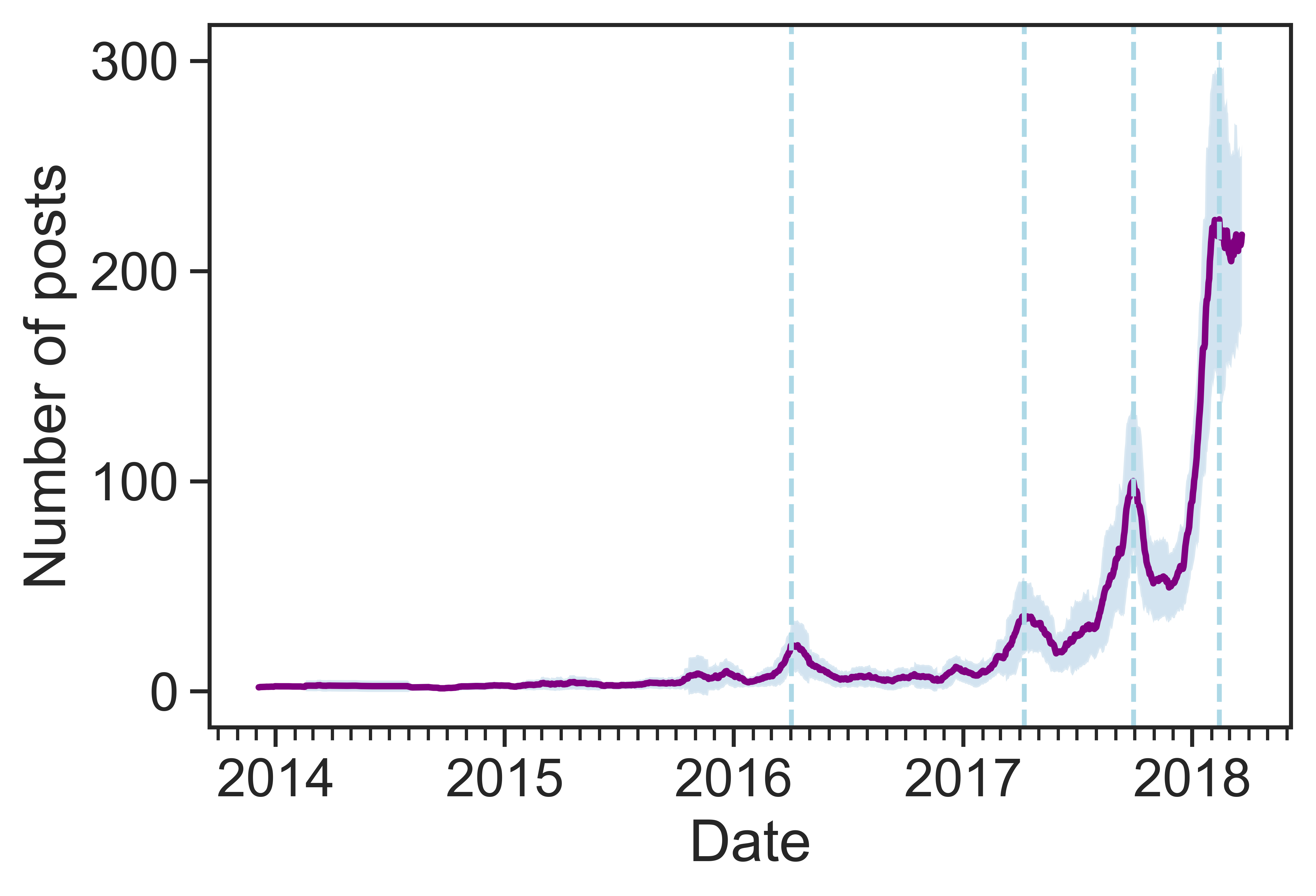}
                \caption{}
                \label{fig:posting_trend}
        \end{subfigure}%
        \begin{subfigure}[b]{0.33\textwidth}
                \includegraphics[width=\linewidth]{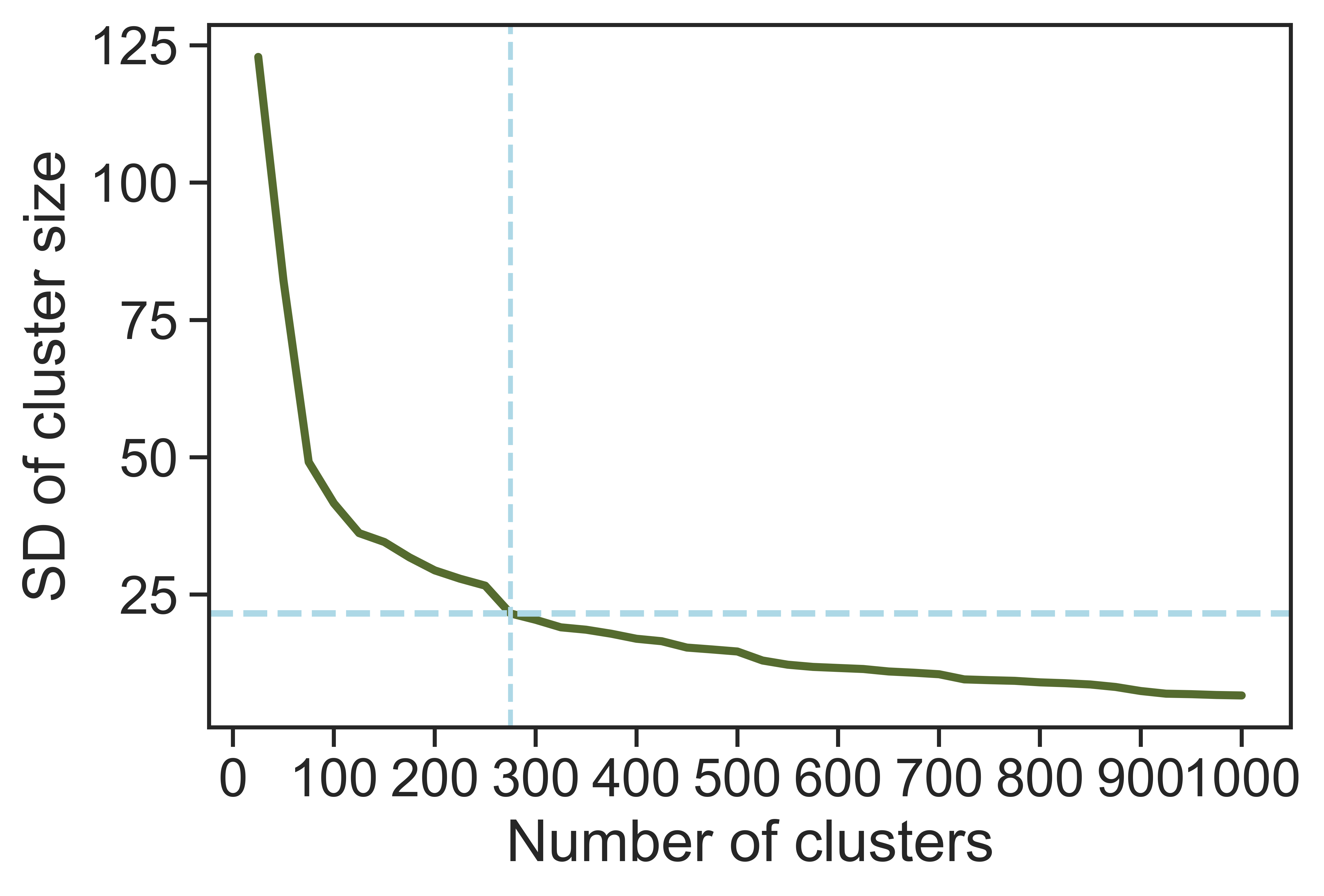}
                \caption{}
                \label{fig:elbow}
        \end{subfigure}%
        \caption{(a) Illustration of the coherence score as a function of the number of topics. (b) The temporal trend of the number of posts in months. (c) Illustration of the standard deviation of cluster size as a function of the number of word semantic clusters.}
\end{figure*}

\subsection{Descriptive Statistics}

We collected 37,183 posts published by 5,857 users between February 28, 2011 and March 21, 2018. Among these posts, there are 3,413 submissions and 33,770 comments. The median (interquartile range) length of submissions is 54 words (22 to 100), while the median (interquartile range) length of comments is 54 words (10 to 43). The median (interquartile range) karma scores received by submissions is 4 (2 to 8), while the median (interquartile range) number of comments received by submissions is 3 (2 to 5). 

\subsection{RQ1: Topics of Discussion}

We first presented the general topics that were communicated in this subreddit through a topic modeling analysis.

Figure~\ref{fig:coherence} illustrates the topic coherence scores changed as a function of the number of topics. We use 12, which corresponds to the highest coherence score, as the optimal number of topics. Table~\ref{tab:topic_terms} shows the inferred topics: their marginal distribution and most relevant terms. The relevance of a term is measured by the probability that it is sampled from a topic. The marginal distribution of a topic is measured by the probability that the topic is sampled from the entire subreddit. We summarized the topics into the following four categories:

\textbf{i) Kinship.} This category mainly corresponds to topic T1, and most words are related to family members and relatives. This topic is about exploring gene similarity with known relatives, but also about kinship search. The following is one illustrative example:

\begin{quote}
``\textit{I was lucky enough to find two kids belonging to my great-grandfather's youngest brother, who I didn't even know was still alive, and his daughter. I made contact with the daughter and have learned so much from talking to her. You never know what you might get!}''
\end{quote}

\textbf{ii) Testing Reports.} This category includes topics T3, T4, T7, T8, and T9, which can be further grouped into three subcategories: 1) ancestry composition (topic T4 and T9); 2) discussion on ancestry (topic T3 and T8); and 3) raw data processing (topic T7). Relevant examples of this category include:

\begin{quote}
``\textit{Results are in! I am 3/4 Appalachian American (Southwest Virginia) and 1/4 New Yorker.}''
\end{quote}

\begin{quote}
``\textit{Hi, I just ran my raw data through Eurogenes K15 V2 and I do NOT understand most of the results, any way you could help me out?}''
\end{quote}

\begin{quote}
``\textit{True my nose is more "African" like I also have curly hair so i get it. I think I get those genes from my mom who isn't completely European.}''
\end{quote}

\textbf{iii) Discussion on Health and Traits.} This category includes topics T2, T5, and T10. Note that T2 is related to disease risk discussion, while topics T5 and T10 are related to general discussion. The following are illustrative examples of this category:

\begin{quote}
``\textit{... grandmother passed from late-onset Alzheimer's, ..., the test told me I have a slightly higher chance of developing it.}''
\end{quote}

\begin{quote}
``\textit{Some study I read said people with the warrior gene have different processing of stress chemicals that make us perform better in a crisis. Since I'm the one who always has to handle emergencies ... I think it's a trait that can be useful.}''
\end{quote}

\begin{quote}
``\textit{This could not be truer ... EVERYONE deserves to feel good about their heritage, and no one deserves to feel bad. This applies to people of all colors and all races.}''
\end{quote}

\textbf{iv) Testing Service Progress:} This category includes topics T6, T11, and T12. T6 is related to communications with companies, while topics T11 and T12 are related to testing timeline. Here are examples:

\begin{quote}
``\textit{I'm beginning to think they knocked over an entire tray of Dec 8th samples}''
\end{quote}

\begin{quote}
``\textit{Received mine November 27th mailed out 28th they received it Dec 2nd and passed inspection. It has been in extraction since ...  I believe it was sent to the facility in North Carolina as well ...}''
\end{quote}

\begin{quote}
``\textit{When I ordered it the expected wait time was 6-8 weeks. They received my sample on Feb 12. It's still on quality review so I'm not sure how much longer it will take.}''
\end{quote}

It should be noted that T1 exhibited the largest distribution in this subreddit, but its category was the smallest.

\begin{table*}[ht]
\centering
\begin{tabularx}{\textwidth}{l|Z|r}
\toprule[1.pt]
ID & \multicolumn{1}{c|}{ Most Relevant Terms}                                                                                                                                                                                                                           & Dist. \\  \midrule[1pt] 
T1       & \textit{father, dad, great, mother, share, side, mom, relative, cousin, match, family, half, parent, grandparent, find, grandmother, grandfather, child, paternal, brother, sister, sibling, relate, maternal, close, biological, adopt, bear, list, story}                        & 9.0\% \\ \hline
T2       & \textit{gene, genetic, https\_www, reddit, risk, comment, variant, snp, link, testing, mutation, snps, high, disease, study, add, trait, http, https, snpedia, promethease, rs, base, http\_www, medical, type, top, information, low, condition}                                  & 8.9\% \\ \hline
T3       & \textit{people, population, group, live, common, country, large, close, neanderthal, area, genetic, https, lot, average, ancestor, part, region, place, base, score, human, due, admixture, single, history, world, genetically, speak, recent, term}                              & 8.7\% \\ \hline
T4       & \textit{european, african, result, italian, ancestry, north, broadly, southern, middle\_eastern, europe, eastern, percentage, jewish, iberian, west, high, ashkenazi, finnish, show, east, northern, noise, balkan, ancestor, percent, category, africa, unassigned, andme, small} & 8.5\% \\ \hline
T5       & \textit{make, thing, find, feel, chromosome, haplogroup, good, question, didn, understand, answer, talk, happen, life, idea, year, sound, sort, point, call, bad, love, doesn, friend, person, real, man, learn, hard, kind}                                                       & 8.4\% \\ \hline
T6       & \textit{sample, time, email, andme, process, lab, work, customer, contact, order, account, issue, send, experience, information, website, response, problem, call, customer\_service, ready, saliva, provide, follow, reply, status, batch, question, company, app}                & 8.4\% \\ \hline
T7       & \textit{andme, report, datum, ancestry, health, give, gedmatch, site, raw\_data, information, good, promethease, chip, service, info, upload, free, run, interested, company, pay, tool, file, worth, buy, myheritage, option, ancestrydna, offer, version}                        & 8.4\% \\ \hline
T8       & \textit{family, back, irish, german, pretty, british, french, guess, lot, expect, find, ve, interesting, bit, white, cool, people, black, ancestor, scandinavian, surprise, tree, generation, english, hair, super, big, kind, line, awesome}                                      & 8.3\% \\ \hline
T9       & \textit{asian, ancestry, result, native\_american, south, east, american, native, andme, chinese, accurate, land, people, https\_imgur, wegene, percentage, mexican, interesting, part, ethnicity, lot, japanese, white, mix, korean, guess, indian, show, true, pretty}           & 8.2\% \\ \hline
T10       & \textit{dna, test, show, delete, change, doesn, give, parent, edit, case, happen, put, person, isn, good, wrong, work, amount, inherit, exact, wouldn, chance, compare, correct, win, assume, reason, composition, actual, recently}                                               & 8.0\% \\ \hline
T11      & \textit{extraction, receive, kit, day, analysis, week, step, send, complete, move, today, mine, fail, report\_generation, jan, computation, processing, timeline, stage, feb, quality\_review, december, batch, january, register, mail, dec, wait, march, husband}                & 7.7\% \\ \hline
T12      & \textit{result, ve, wait, long, time, post, check, update, week, hope, phase, month, start, ago, hear, mine, hop, hour, morning, today, haven, end, didn, couple, read, anyone\_else, lol, finally, thread, nice}                                                                  & 7.6\% \\ \midrule[1pt] 
\end{tabularx}
\caption{The 12 topics that were identified from \textit{r/23andme}. The sample words are ordered based on their relevance to the topic. The distribution of each topic is calculated on the posts from the entire subreddit.}
\label{tab:topic_terms}
\end{table*}

\subsection{RQ2: Temporal Trends}
Next, we investigated the topic temporal trends and how they were related to the posting temporal trends. 

\textbf{Posting Temporal Trends.} Figure~\ref{fig:posting_trend} illustrates how the number of posts per month changed over the time. From the figure, it can be seen that: 1) Initially, the number of posts remained very low, and the trend was almost flat before 2017. However, there was a rapid increase in 2017. For example, the number of posts around 2018 had increased to approximately 200 per month, while at the beginning of 2017, this number was only approximately 10; 2) There are four date months (from left to right) that corresponded to the local maxima of the number of posts: March 2016, April 2017, September 2017, and February 2018.

\begin{figure*}[ht]
\centering
  \includegraphics[width=2.0\columnwidth]{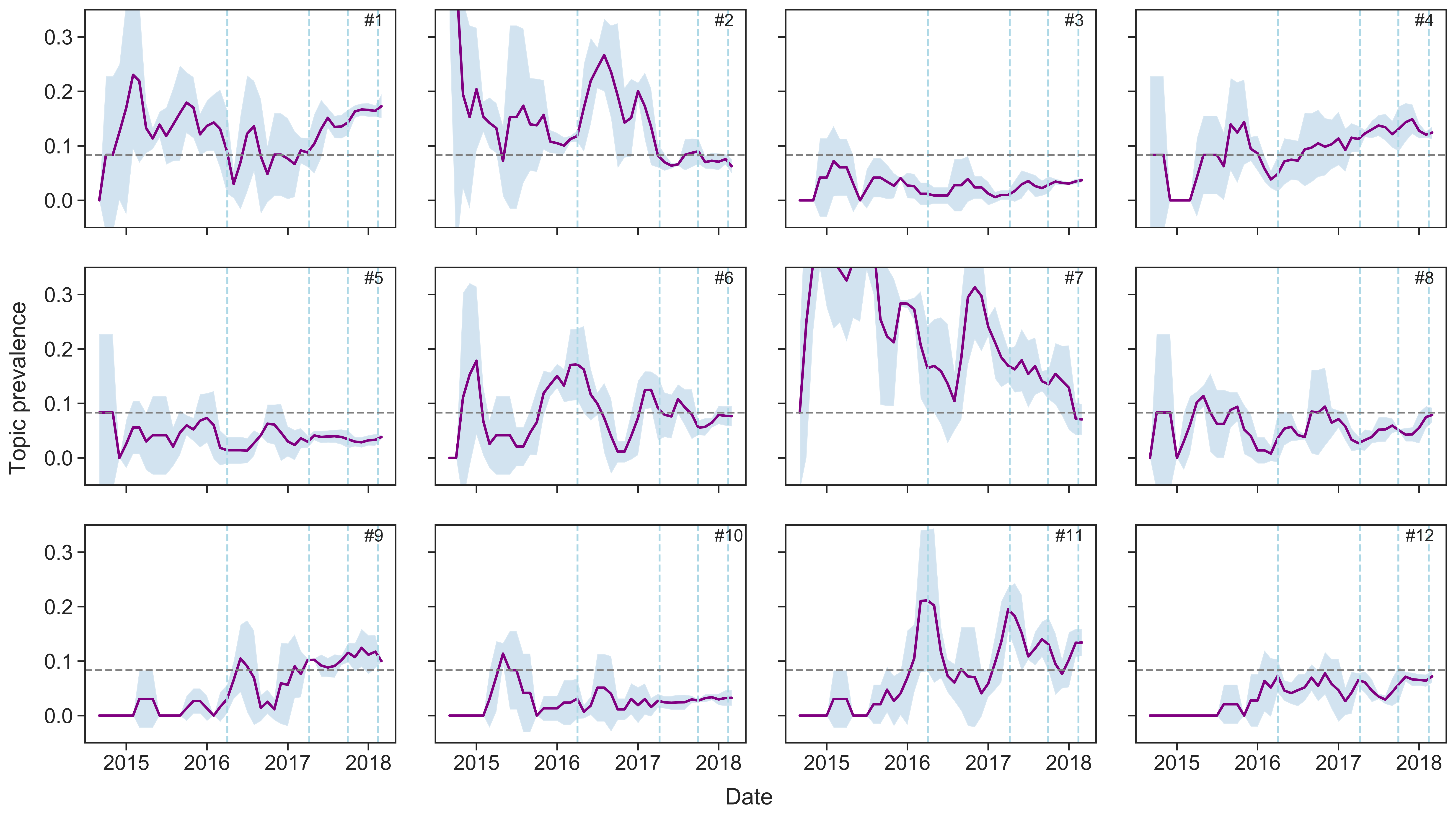}
  \caption{Temporal trends of the topic prevalence. The horizontal dashed line corresponds to the average prevalence of 8.3\% if topics are evenly distributed. The four vertical dashed line are corresponding to the dashed lines shown in Figure~\ref{fig:posting_trend}.}~\label{fig:topic_prevalence_trend}
\end{figure*}

\textbf{Topic Temporal Trends.} Figure~\ref{fig:topic_prevalence_trend} illustrates how the prevalence of the topics of submissions changed over time. The horizontal dashed line corresponds to the average prevalence level ($8.3\% \approx 1.0/12$), under the assumption that all topics were equally distributed. The purple lines above this dashed line suggest an above average prevalence, while the purple lines below it suggest a below average prevalence. The vertical dashed lines correspond to the four date months of the local maxima in Figure~\ref{fig:posting_trend}.

There were several findings worth noting. First, due in part to the limited number of posts before March 2017 (see the second vertical dashed line in Figure~\ref{fig:posting_trend}), the prevalence of all the topics had a relatively large variance before this time period.

Second, before March 2017, topics T1, T2 and T7 were the three most prevalent topics. However, after March 2017, topic T1 rose above the average prevalence level, while T2 was a little bit lower than the average level, and T7 dropped to the average level. Additionally, after March 2017, T4 and T9 were slightly above the average level and exhibited a gently increasing trend. By contrast, T11 was above the average, but experienced a decreasing trend before 2018, and then an increasing trend after 2018.

Third, topics T3, T5, T8, T10 and T12 were usually less frequently discussed than the other topics, especially after March 2017.

Fourth, T11 achieved local maxima at all the four date months indicated at the vertical dashed lines, followed by T12 which had its local maxima at the three earlier date months, and then T6 which had local maxima at the first date month. Note that all the three topics were related to topic category iv, testing service progress.  

\begin{table*}[ht]
\centering
\begin{tabularx}{\textwidth}{l|Z|l|l|l|r}
\toprule[1.pt]
ID &  \multicolumn{1}{c|}{Sample Words}                                                                                                                                                                                                   & $\beta$ & Z    & Std   & Pvals     \\ \midrule[1pt] 
C74 & \textit{disease, cancer, diabetes, lupus, cancers, glaucoma, diseases, thyroid, diagnosis, melanoma, kidney, liver, infection, celiac, disorders, allergies, diagnosed, prostate, autism, symptoms} & 1.276 & 5.369 & 0.238 & $7.9\times 10^{-8}$ \\ \hline
C230 & \textit{established, developed, develop, formed, establish, acted, served, implemented, adopted, developing, serve, serves, modeled, launched} & 1.271 & 4.353 & 0.292 & $1.3\times 10^{-5}$ \\ \hline
C5 & \textit{reach, reaching, reached, settled, reaches, exceeded, cleared, exceed, meet, settle, settling, met, achieve, achieved, approaching, clearing, meets, blocked, settlement, sealed} & 1.169 & 3.921 & 0.298 & $8.8\times 10^{-5}$ \\ \hline
C213 & \textit{pregnancy, babies, newborn, baby, pregnant, fetus, fertility, infertile, miscarriages, miscarriage, sperm, maternal, infant, ovaries, inseminated, child, reproductive, insemination, ivf, ovarian} & 1.073 & 4.972 & 0.216 & $6.6\times 10^{-7}$ \\ \hline
C225 & \textit{parents, children, adoptive, kids, fathers, foster, adoptees, grandparents, adoptee, parental, dads, counselors, counselor, parent, adoptions, therapist, counseling, psychiatrist, orphan, adoption} & 0.978 & 4.992 & 0.196 & $6.0\times 10^{-7}$ \\ \hline
C151 & \textit{mother, son, daughter, father, niece, aunt, grandmother, uncle, dad, cousin, brother, nephew, sister, granddaughter, husband, daughters, grandson, siblings, wife, sons} & 0.976 & 6.565 & 0.149 & $5.2\times 10^{-11}$ \\ \hline
C160 & \textit{medical, physician, patients, patient, doctors, doctor, treatment, hospital, inpatient, health, clinic, medicine, treatments, clinical, healthcare, diagnostic, therapy, neurologist, detoxification, wellness} & 0.865 & 3.985 & 0.217 & $6.7\times 10^{-5}$ \\ \hline
C224 & \textit{decades, centuries, century, decade, millenia, era, modern, years, neanderthals, revolution, caveman, neanderthal, revolutionary, remnants, legacy, paleo, distantly, distant} & 0.811 & 4.101 & 0.198 & $4.1\times 10^{-5}$ \\ \hline
C69 & \textit{thats, dont, america, ethiopia, tho, penn, ellis, mcdonald, maryland, tennessee, american, doug, patrick, norman, im, david, oklahoma, dominican, alaska, george} & 0.798 & 5.115 & 0.156 & $3.1\times 10^{-7}$ \\ \hline
C145 & \textit{week, morning, day, afternoon, evening, month, night, weekend, noon, hours, 5pm, days, midnight, 7pm, 8am, hour, weeks, tomorrow} & 0.684 & 4.405 & 0.155 & $1.1\times 10^{-5}$ \\ \hline
C198 & \textit{statistics, numbers, percentages, statistic, stats, figures, statistically, statistical, percentage, records, number, results, proportion, outcomes, record, histories, history, outcome} & 0.663 & 4.164 & 0.159 & $3.1\times 10^{-5}$ \\
\midrule[1pt] 
\end{tabularx}
\caption{Word semantic clusters that were statistically significant with respect to karma score received by a submission. The sample words are ordered based on their distances to the cluster centers.}
\label{tab:karma_clusters}
\end{table*}

\subsection{RQ3: Attention to Submissions}
We further examined the types of content that were more likely to receive high attention, as defined by karma score and the number of comments.

Figure~\ref{fig:elbow} shows how the standard deviation of cluster size changed as a function of the number of clusters. We set the number of word semantic clusters to 275. This was done because, after this point, the decrease of standard deviation of cluster size flattened out. After removing correlated variables, there were 9 topics, 40 linguistic features and 271 word semantic clusters that were applied to fit models.

\textbf{Karma Score.} Our analysis showed that post length and linguistic feature \textit{conjunctions} exhibited a negative association with the karma score received by a submission ($\beta$=-1.940, P=$4.3\times10^{-9}$; $\beta$=-0.784, P=$3.7\times10^{-6}$); Topic T5 and linguistic feature \textit{2nd person} had a positive association with karma score ($\beta$=1.646, P=$2.3\times 10^{-6}$; $\beta$=1.188, P=$1.4\times 10^{-8}$). Table~\ref{tab:karma_clusters} shows the word semantic clusters that were statistically significant with respect to  karma score. 

There were several observations worth noting. First, all of the clusters achieved a statistically positive association with karma score. Among these features, the cluster related to health issues (C74, $\beta$=1.276) exhibited the largest association with karma score. Additionally, mentions of other health related words (C160, $\beta$=0.865) achieved a positive association with karma score as well. A notable example is shown as follows:

\begin{quote}
``\textit{23andme potentially saved my life. ... I've been seeing doctors for the past 6+ years ... Nobody could pin down a specific reason ... 23andme found rare genetic mutations that cause HFE-related Hereditary Hemochromatosis and I was diagnosed less than 1 week later.}''
\end{quote} 

Second, mentions of words related to kinship, either gene match between biologic relatives or kinship search (C213, $\beta$=1.073; C225, $\beta$=0.978; C151, $\beta$=0.976), were statistically significant with respect to karma score. Illustrative examples of this finding were:

\begin{quote}
``\textit{... Got the test results back from our cousin and neither one of us is related to him ... Time to talk to mom ... (UPDATE) My dad was exposed to radiation ... and became infertile ...}''
\end{quote}

\begin{quote}
``\textit{... I am in shock. I found my birth family... after almost 50 years ...  I was given up for adoption in a closed adoption meaning I have few ways short of a court order of finding out who my birth family is ...} ''
\end{quote}

Finally, other statistically significant clusters included mentions of action words (C230 $\beta$=1.271; C5, $\beta$=1.169), ancestry related discussion (C224, $\beta$=0.811), locations (C69, $\beta$=0.798), and time or statistics related words (C145, $\beta$=0.684; C198, $\beta$=0.663).

\begin{table*}[ht]
\centering
\begin{tabularx}{\textwidth}{l|Z|l|l|l|r}
\toprule[1.pt]
ID &  \multicolumn{1}{c|}{Sample Words}                                                                                                                                                                                                   & $\beta$ & Z    & Std   & Pvals     \\ \midrule[1pt] 
C236 & \textit{script, edited, scripts, editing, edit, creators, creator, moderator, cartoon, journalist, commenter} & 1.038 & 5.900 & 0.176 & $3.6\times 10^{-9}$ \\ \hline
C224 & \textit{decades, centuries, century, decade, millenia, era, modern, years, neanderthals, revolution, caveman, neanderthal, revolutionary, remnants, legacy, paleo, distantly, distant} & 0.655 & 4.819 & 0.136 & $1.4\times 10^{-6}$ \\ \hline
C128 & \textit{methods, extraction, method, techniques, mixtures, algorithms, processes, methodology, extract, admixtures, extracting, algorithm, extracted, derived, mathematical, admixture, quantum, extractions} & 0.650 & 6.449 & 0.101 & $1.1\times 10^{-10}$ \\ \hline
C59 & \textit{calculate, computed, calculated, calculation, calculating, calculations, compute, decimal, equals, computation, average, recalculated, minimum, maximum, equivalent, max, plus, minus, equal, averages} & 0.579 & 5.000 & 0.116 & $5.7\times 10^{-7}$ \\ \hline
C145 & \textit{week, morning, day, afternoon, evening, month, night, weekend, noon, hours, 5pm, days, midnight, 7pm, 8am, hour, weeks, tomorrow} & 0.432 & 4.136 & 0.104 & $3.5\times 10^{-5}$ \\ \hline
C72 & \textit{3x, 6x, 7x, 4x, 2x, 5x, ratio, volume, ratios, downside, upside, peak, usd, peaks, volumes, chart, rate, rates, ranges, charts} & -0.897 & -4.012 & 0.224 & $6.0 \times 10^{-5}$ \\
\midrule[1pt] 
\end{tabularx}
\caption{Word semantic clusters that were statistically significant with respect to the number of comments received by a submission. The sample words are ordered based on their distances to the cluster centers.}
\label{tab:comment_clusters}
\end{table*}

\textbf{Number of Comments.} We found that the number of days since a submission was published was negatively associated with the number of comments (coeff=-0.459, P=$4.8 \times 10^{-10}$); the linguistic feature \textit{future focus} and topic T7 were negatively associated with the number of comments (coeff=-0.783, P=$9.4\times 10^{-5}$; coeff=-0.875, P=$7.6\times 10^{-5}$); the linguistic features \textit{sad}, \textit{2nd person} and \textit{tentative} were positively associated with the number of comments (coeff=1.079, P=$0.6\times 10^{-5}$; coeff=0.920, P=$2.2\times 10^{-10}$; 0.783, P=$1.5\times 10^{-5}$). Table~\ref{tab:comment_clusters} shows the word semantic clusters that were statistically significant with respect to the number of comments. 

There were several notable observations from this investigation. First, the cluster with words related to \textit{editing} had the largest positive association with the number of comments (C236, $\beta$=1.038). A supportive example of this finding is:

\begin{quote}
``\textit{... journalist seeking 23andme user ... I am interested in how you reacted to your results, and if you acted on these results ...}''
\end{quote}

Second, discussion about the testing methods (C128, $\beta$=0.650; C59, $\beta$=0.579), or timeline related to obtain testing results (C145, $\beta$=0.432) were positively associated with the number of comments. An example of such discussion is:

\begin{quote}
``\textit{Am I the only one who has been stuck in extracting for over 4 weeks? }''
\end{quote}

Third, mentions of ancestry related words (C224, $\beta$=0.655) were also positively associated with the number of comments. However, mentions of ratio or rate related words (C72, $\beta$=-0.897) was negatively associated with the number of comments. An example of this observation is:

\begin{quote}
``\textit{... 23andme tells me I have a very high chance of NOT going bald before the age of 40. However, Promethease says I have a 7x more likely chance of baldness ...}''
\end{quote}

\section{Discussion}
\label{sec:discussion}

In this section, we summarize our primary findings, and discuss the implications, limitations and future work.

\subsection{Online Discussion Topics}

We identified 12 topics that could be further grouped into four main categories: i) kinship, ii) testing reports, iii) discussion of health and traits, and iv) testing service progress. Among these topics, T1, which corresponds to biological relatives, exhibited the largest distribution (9.0\%). This was followed by topic T2 (8.9\%), which corresponds to discussion about health risks. Notably, this illustrates how Reddit differs from Twitter where people are more likely to share their ethnicity results \cite{olejnik2014m}, This further suggests that \textit{r/23andme} users disclosed richer content regarding DTC genetic testing. 

It should be noted that topic T2 also included the names \textit{Snpedia} and \textit{Promethease}, which are popular third-party platforms for interpreting genetic testing results, especially about disease risks. This is particularly interesting because it implies two potential important issues that were investigated frequently by previous studies (see Background and Motivations section): First, how is the genetic information of users protected when it is shared for third-party interpretation? Who is responsible in the event that a privacy intrusion transpires? Second, to what extent should users trust the interpretation or evaluation of disease risks \cite{moscarello2018direct}? When there is a conflict in the interpretation between two parties, who will help explain the discrepancy to consumers? 

Another popular topic that was discussed was about ethnicity composition, which is understandable because it is a typical task of 23andMe genetic testing service. Similarly, it was not surprising to observe that checking the progress of applications was another important subject for people to discuss in this subreddit. 

\subsection{Temporal Trends}

We found that the posting trend was aligned with the distribution of the total number of users who have already taken DTC genetic testing reported in \cite{regalado2018}. Both experienced a substantial increase during 2017. 

When examining the topic trends at the four date months where the number of posts achieved local maxima, we found that discussion about testing progress (Topics T11, T12 and T6) also achieved local maxima. This suggests that many users purchased the service from 23andMe during or before these data months. Digging into the matter a bit deeper, we found that some of coincidences could align to two events regarding 23andMe. First, in April 2017 (around the data month at the second vertical dashed line in Figure~\ref{fig:posting_trend}), FDA granted market authorization to 23andMe for reporting the health risk for ten diseases \cite{FDA2017}. It is plausible that this news drove more people to purchase the service. Second, it was reported that the 23andMe genetic testing kit was one of Amazon's five best-selling items on Black Friday in 2017 \cite{estrada2017}. This may explain why there was a huge jump on the number of posts at the beginning of 2018 (around date month at the fourth vertical dashed line in Figure~\ref{fig:posting_trend}). For example, as a user mentioned in a post:

\begin{quote}
``\textit{... I order mine on Black Friday on Amazon for a total of \$106. And got my results in January. It was the ancestry+health ...}''
\end{quote}

The topic trend analysis indicated that kinship was steadily growing towards a major topic that people shared or sought suggestions in \textit{r/23andme}. However, it should be noted that there are two different types of behaviors regarding kinship. The first is about gene matching between known family members, while the second is about kinship search for the unknown relatives. Both behaviors may lead to unexpected results. As shown in earlier quotes, some people became excited because they found their biological parents, while others became frustrated because they found that their fathers were not their biological ones. No matter which results were realized, these are the real consequences that people have to face once they open the Pandora's box.

The topic related to the discussion on raw data processing offered by different companies (T7) had experienced a steady decrease in 2017. Considering that there were mainly 23andMe customers in this subreddit, and as more people adopted this service, the number of submissions about this topic might not increase as quickly as other topics. Despite a decreasing trend, it was still a major topic that people want to ask in this subreddit. The prevalence of discussion on ancestry composition was slightly growing above the average (topics T4 and T9), again, might be due to the fact that ancestry service is a basic product offered by 23andMe.   

\subsection{Association With Submission Attention}

We found that the length of a post had a negative association with karma score, while the live time had a negative association with the number of comments. Based on Table~\ref{tab:karma_clusters}, we can approximately summarize the word semantic clusters in a descending order by their likelihoods to associate with high karma scores: health (C74 and C160); kinship (C213, C225 and C151); ancestry composition (C224, C69 and C198); and timeline to obtain results (C145). While the prevalence of the disease risk related topic (T2) was slightly lower than the average in 2017 (see Figure~\ref{fig:topic_prevalence_trend}), health related word semantic clusters were most likely to obtain high karma score.  It should be noted that among the kinship clusters, the adoption related semantic cluster (C225) is also more likely to receive high karma score. By contrast, from Table~\ref{tab:comment_clusters}, it can be seen that submissions discussing testing reports or ancestry were more likely to receive comments. We suspect the differences might be caused by the different motivations of making comments and making votes. However, the related investigation is beyond the current research.  

Finally, from a methodological perspective, we applied a combination of topics, linguistic features and word semantic clusters to obtain a rich profile of the discussion content. It should be noted that many topic and linguistic features were not statistically significant due to the incorporation of word semantic clusters into our models, suggesting the importance of word clusters in explaining the attention received by submissions. However, it was still observed that submissions communicating a sad emotion were more likely to receive comments, which was similar to \cite{de2014mental} where negative emotion was positively associated with the number of comments. 

\subsection{Implications}
Our study has two main implications. First, health and kinship (including kinship search) were discussed and received substantial attention in this subreddit, but the consequences of doing so were mixed in that some people found the results to be positive while others appeared to be disturbed. 

Second, undergoing the genetic test or seeking for third-party interpretation may lead to privacy intrusions with respect to one's genetic information. While there were posts regarding such concerns, our analysis did not find strong signals across this subreddit. Yet, the interesting quote from a submission might provide some intuition into some users' attitude about this topic:

\begin{quote}
``\textit{How shady is 23andme? Should I just avoid the DNA testing if I value my privacy? ... articles like this and the past conduct of similar data-gathering companies like Google has me worried ... EDIT: And if someone could explain why I'm being downvoted just for asking privacy questions, that'd be great.}''
\end{quote}

\subsection{Limitations and Future Work}
There are several limitations that we believe can serve as the basis of future work. First, the population of our study was composed of active users in \textit{r/23andme}, which may limit the generalizability of our findings. It will be interesting to take into consideration of users from other subreddits (e.g. \textit{r/AncestryDNA}) or other online platforms. Second, we relied on karma score and the number of comments to capture the attention that a submission received. However, it is not straightforward to explain the motivation for Reddit users to comment or vote a post. Future work may consider to develop a better metric to evaluate people's attention to the discussion of DTC genetic testing. We applied three different types of features to characterize the post content, and focused on providing a general picture on discussion in this subreddit. It will be worthwhile to investigate to what extent the online discussion could help individuals cope with the consequences of undergoing DTC genetic testing.
\section{Conclusion}
\label{sec:conclusion}
In this study, we investigated online discussion of direct-to-consumer genetic testing in \textit{r/23andme}. Specially, we first applied topic modeling to extract topics that were discussed in this subreddit, and then studied topic and posting temporal trends. We further applied regression analysis to learn the association between the attention that a submission received, in terms of karma score and the number of comments it received, and the post content, in terms of topics, linguistic features and word semantic clusters. We found that there was a rapid increase in the online discussion during 2017. While ancestry composition was a popular subject, kinship was steadily growing towards a major online discussion topic. Moreover, compared with other topics, health and kinship were more likely to receive attentions in terms of votes, while testing reports were more likely to receive attentions in term of comments. Our findings are further evidence that people may not always be prepared to deal with the consequences of DTC genetic tests, and the users in this particular subreddit might not consider privacy sufficiently when taking the test or request third-party interpretation. 
\section{Acknowledgments}
This work was supported by the National Science Foundation grant number IIS1418504. 

\begin{small}
\bibliography{Bibliography}
\bibliographystyle{aaai}
\end{small}

\end{document}